\documentclass[aps,twocolumn,superscriptaddress,amsfont,graphicx,nofootinbib,preprintnumbers]{revtex4-1}%
\UseRawInputEncoding
\usepackage{color,graphicx,epsfig}
\usepackage{xcolor}
\usepackage{ifpdf}
\usepackage{amsmath}
\usepackage{bm}
\usepackage[english]{babel}
\usepackage{multirow}
\usepackage{amssymb}
\usepackage{braket}
\usepackage{hyperref}
\usepackage{enumerate}
\usepackage{url}
\usepackage{slashed}
\usepackage[varg]{txfonts}   

\def\({\left(}
\def\){\right)}

\def\beq{\begin{equation}}
\def\eeq{\end{equation}}
\begin{document}

\title{
Footprints of Axion-Like Particle in Pulsar Timing Array Data and James Webb Space Telescope Observations
}

\author{Shu-Yuan Guo}
\email{shyuanguo@ytu.edu.cn}
\affiliation{Department of Physics, Yantai University, Yantai 264005, China}

\author{Maxim Khlopov}
\email{khlopov@apc.univ-paris7.fr}
\affiliation{Institute of Physics, Southern Federal University, Stachki 194 Rostov on Don 344090, Russia}
\affiliation{National Research Nuclear University MEPhI (Moscow Engineering Physics Institute), 115409
Moscow, Russia}
\affiliation{Virtual Institute of Astroparticle physics, rue Garreau, 75018 Paris, France}

\author{Xuewen Liu}
\email{xuewenliu@ytu.edu.cn}
\affiliation{Department of Physics, Yantai University, Yantai 264005, China}

\author{Lei Wu}
\email{leiwu@njnu.edu.cn}
\affiliation{Department of Physics and Institute of Theoretical Physics, Nanjing Normal University, Nanjing, 210023, China}

\author{Yongcheng Wu}
\email{ycwu@njnu.edu.cn}
\affiliation{Department of Physics and Institute of Theoretical Physics, Nanjing Normal University, Nanjing, 210023, China}

\author{Bin Zhu}
\email{zhubin@mail.nankai.edu.cn}
\affiliation{Department of Physics, Yantai University, Yantai 264005, China}

\begin{abstract}
Several Pulsar Timing Array (PTA) collaborations have recently reported the evidence for a stochastic gravitational-wave background (SGWB), which can unveil the formation of primordial seeds of inhomogeneities in the early universe. With the SGWB parameters inferred from PTAs data, we can make a prediction of the seeds for early galaxy formation from the domain walls in the axion-like particles (ALPs) field distribution. This also naturally provides a solution to the observation of high redshifts by the James Webb Space Telescope. The predicted photon coupling of the ALP is within the reach of future experimental searches.

\end{abstract}
\maketitle
\section{Introduction}
\label{intro}

The groundbreaking discovery of gravitational waves (GWs) by the LIGO-Virgo collaboration~\cite{LIGOScientific:2016aoc, LIGOScientific:2016sjg, LIGOScientific:2017bnn, LIGOScientific:2017vox, LIGOScientific:2017ycc, LIGOScientific:2017vwq, LIGOScientific:2020zkf, LIGOScientific:2020iuh} has opened up an extraordinary pathway for our understanding of the universe. Unlike light waves, GWs allow us to probe the universe in its infancy. One of the main ways by which GWs can unveil the early universe comes from a stochastic GW background. The very recent analyses by Chinese Pulsar Timing Array (PTA)~\cite{Xu:2023wog}, European PTA~\cite{Antoniadis:2023ott}, NANOGrav~\cite{NANOGrav:2023gor}, Parkes PTA~\cite{Reardon:2023gzh} have reported the compelling evidence for such a stochastic common-spectrum process, which can be produced in a variety of astrophysical and cosmological processes~\cite{NANOGrav:2023hvm, Madge:2023cak, Bian:2020urb, Bian:2023dnv}, such as the supermassive black hole binaries~\cite{NANOGrav:2023hfp}, the phase transitions~\cite{PhysRevD.25.2074, Kosowsky:1991ua, Cai:2017tmh, Addazi:2023jvg, Xue:2021gyq, Deng:2023seh}, preheating events~\cite{Khlebnikov:1997di, Liu:2017hua}, the presence of topological defects~\cite{PhysRevD.31.3052, Auclair:2019wcv, Saikawa:2017hiv, Zhou:2020ojf, Sakharov:2021dim, King:2023cgv, Chiang:2020aui, Bian:2022tju}, and the existence of high-amplitude scalar perturbations~\cite{Ananda:2006af, Baumann:2007zm, Kohri:2018awv}. On the other hand, a powerful infrared telescope, the James Webb Space Telescope (JWST) has observed, for the first time, the existence of primordial galaxies and early Active Galactic Nuclei (AGN) formation at high redshifts~\cite{Labb__2023}. The presence of the high redshift galaxies requires an unexpectedly high star formation efficiency and thus challenges the standard cosmological model $\Lambda$CDM. All these findings provide invaluable insights into the primordial structure formation and the new physics beyond the Standard Model.

In this work, we proposed a new mechanism of the closed axion domain wall as the common explanation of the SGWB detected by PTAs and the high redshift AGN observed by JWST. Axion-like particle (ALP) is a well-motivated extension of the Standard Model that could play a significant role in elucidating the unknown aspects of cosmology and particle physics. Addressing the strong CP problem in quantum chromodynamics (QCD)~\cite{Peccei:1977hh} or considering compactification within string theory~\cite{Arvanitaki:2009fg, Svrcek:2006yi, Conlon:2006tq} may involve the Peccei-Quinn (PQ) or PQ-like symmetries, leading to the emergence of various axions and axion-like particles. The precise energy scale associated with this symmetry remains uncertain, which can be either prior to or during the inflationary phase~\cite{Arvanitaki:2009fg}. The non-perturbative potentials of these models contain terms that cause a continuous degeneracy of the asymmetric ground state. {The manifest symmetry breaking assigns a vacuum expectation value and a phase to the ALP field during inflation. Post-inflation, phase fluctuations and $\pi$ crossings shape domain walls and could lead to the SGWB.}
The destiny of the domain walls after the inflation depends on their radius. If a domain wall has a radius below a critical threshold, it collapses due to its tension. It descends within the cosmological horizon early, thereby giving rise to the creation of primordial black holes (PBHs)~\cite{Rubin:2000dq, Khlopov:2004sc}. On the other hand, if a wall exceeds the critical threshold, its repulsive gravitational field takes over long before it approaches the cosmological horizon, resulting in a wormhole~\cite{Garriga:2015fdk, Deng:2016vzb}. GWs would be generated as the response from the radiation fluid on the repulsion. The repeated processes at different e-folding can lead to the spontaneous formation of both primordial black holes and wormholes. Their evolution can generate a distinct spectrum of gravitational waves, which are indicative of the PQ-like phase transition before or during inflation leading to the closed axion domain wall formation. On the other hand, the contours of these walls are surrounded by regions of the enhanced energy density of the ALP field, which can act as seeds for the early formation of massive galaxies, e.g. through accretion processes~\cite{Liu:2022bvr, Yuan:2023bvh}, and can explain the existence of galaxies formed at high redshifts observed by JWST. Finally, a joint explanation of PTAs and JWST results predicts the ALP mass that can be tested in future experimental searches.

\section{Gravitational Waves and Primordial Structures from Axion-like particle physics}

Axion-like particle (ALP) models can be described by a complex field $\Phi = \phi \exp(i \theta)$ with a spontaneously broken global $U(1)$ symmetry. The potential $V = V_0 + \delta V$ consists of two terms. The first term,
\begin{equation}
V_0 = \frac{\lambda}{2} (\Phi^* \Phi - F_a^2)^2 \label{vo},
\end{equation}
induces spontaneous symmetry breaking of the $U(1)$ symmetry, resulting in a continuously degenerate ground state given by $\Phi_{vac} = F_a \exp(i \theta)$ \cite{book2, PPNP}. The second term,
\begin{equation}
\delta V(\theta) = \Lambda^4 (1 - \cos\theta) \label{disc},
\end{equation}
with $\Lambda \ll F_a$, explicitly breaks the residual symmetry and leads to a discrete set of degenerate ground states, characterized by $\theta_{vac} = 0, 2\pi, 4\pi, \ldots$. The presence of the term (\ref{disc}), either initially or generated by instanton transitions, is a distinct feature of ALP models. The symmetry breaking induced by this term would generate a pseudo-Nambu-Goldstone field $a = F_a \theta$, which acquires a mass $m_{a} = \Lambda^2/F_a$.

\begin{figure}[!htbp]
\begin{center}
\includegraphics[width=8cm]{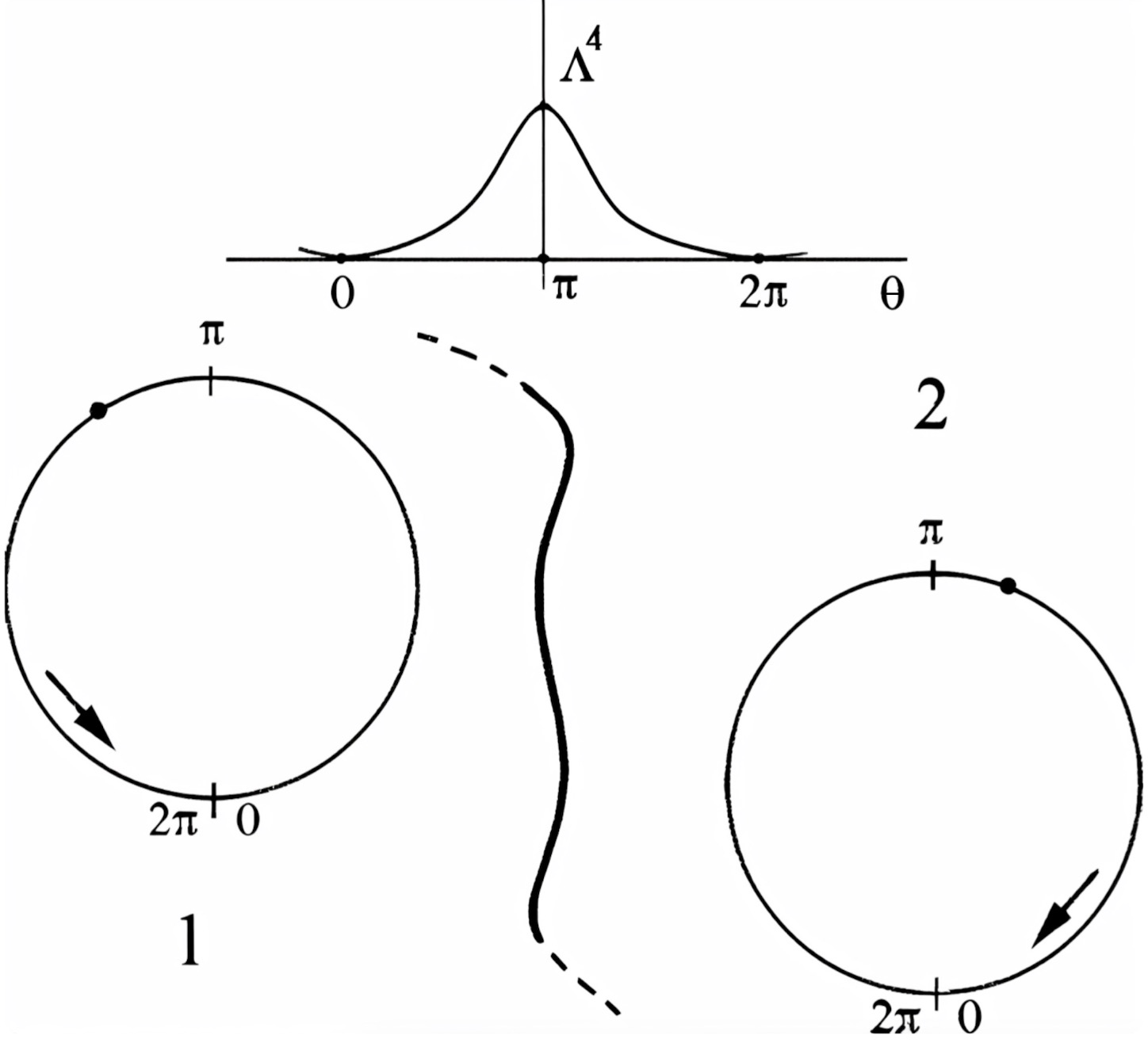}
\caption{Demonstration for phase $\theta$ fluctuation and $\pi$ crossing, crossing of $\pi$ determines the contours of future domain walls. See the main text for details.}
\label{fig:walls}
\end{center}
\end{figure}

Assuming the $U(1)$ symmetry is broken before or during inflation, the field $a$ begins with a misalignment angle $\theta_0$. During inflation, the dynamics of $a$ are influenced by the fluctuation determined by $F_a$ and the Hubble parameter $H_{\rm infl}$. The large Hubble friction during inflation freezes the fluctuation in each e-fold, resembling a one-dimensional Brownian motion for $\theta$ with a step size of $\delta\theta_e = H_{\rm infl}/(2\pi F_a)$ and a typical wavelength of $H_{\rm infl}^{-1}$. With each e-folding, the phase domain splits into $e^3$ disconnected domains. Each of them has a radius of $H_{\rm infl}^{-1}$. {At successive stages of inflation, which define the initial conditions for the regions much smaller than the modern cosmological horizon, the phase fluctuates and can cross $\pi$ at smaller scales. Crossing of $\pi$ at the $i$-th step of inflation determines the contours of future domain walls, while the approach to $\pi$ at the preceding $i+m$ steps ($m=1,2,..$) of inflation gives rise to the local values of $\theta$ and to the corresponding local enhancement of the ALP energy density. Fig.~\ref{fig:walls} is a demonstration for this process.}
The probability that represents the likelihood of a fluctuating phase crossing $\pi$ for the first time at a specific e-fold value $N$ is given as~\cite{Sakharov:2021dim}
\begin{equation}
p\left(\delta \theta_\pi, N\right)=\left(\frac{1}{N}\right)^{3 / 2} \frac{1}{\sqrt{2\pi}} \left(\frac{\delta\theta_\pi}{\delta\theta_e}\right)^2  e^{\frac{-1}{2N} \left(\frac{\delta \theta_\pi}{\delta\theta_e}\right)^2}.
\label{equ:random_walk}
\end{equation}
Here, we define $\delta\theta_{\pi}=|\pi-\theta_0|$ as the initial separation, $\theta_0$ denotes the misalignment angle that field $a$ begins with. This scenario notably differs from other axion models as the formation of domain walls occurs within the inflation period. {Furthermore, such structures appear in the case of the potential $\delta V=\Lambda^4(1+\cos{N\theta})$ with $N=1$, while traditionally axion domain walls are considered in axion cosmology for the case $N>1$~\cite{LAZARIDES198221,Dine:2023qsq}.} The crossings, which can cluster and expand during subsequent stages of inflation, trace the contours where domain walls form at a later time. The formation of the ALP closed domain wall network originating from the inflationary period plays a pivotal role in shaping the primordial structure of the universe. Specifically, it leads to the generation of gravitational waves and the emergence of enhanced ALP energy density regions surrounding contours of closed walls.

After the inflationary period, during the Friedmann-Robertson-Walker (FRW) expansion, the evolution of closed domain walls is size-dependent and can result in their collapse, leading to the formation of primordial black holes (PBHs), or their escape into baby universes, leaving behind wormholes in our universe. Both of these cosmic processes emit GWs, contributing to the stochastic background of GWs. The closed walls, which dominate the mass over the energy of the interior fluid at the Hubble radius, possess a repulsive nature that pushes away the radiation bulk fluid from their interior surface, causing super-luminary expansion and the formation of wormholes to baby universes. In response to this repulsion, the radiation fluid produces sound waves with a characteristic wavelength of approximately $H^{-1}$, setting in motion a fraction of the mass contained inside the Hubble radius and generating stochastic GWs with a frequency of approximately $f\simeq H$. For small walls, the stretching of folded segments on the domain walls, comparable in scale to the Hubble horizon, results in oscillations on a scale similar to the entire closed domain wall, which can be damped, releasing their energy in the form of GWs.

The abundance of the stochastic GW background generated by the closed domain walls can be expressed as~\cite{Sakharov:2021dim}
\begin{equation}
\label{equ:GW}
\Omega_{\rm GW}(f) = \frac{8\pi\sqrt{2}\Omega_r^{1/4}}{135 H_0^{3/2}} \cdot \Gamma_s \cdot
\begin{cases}
    \frac{1}{3}\kappa^2f^{3/2}\\
    \frac{4}{7}(G\sigma)^2f^{-1/2},
\end{cases}
\end{equation}
where the radiation density of the universe $\Omega_r=9.1476\times 10^{-5}$. $\sigma=8m_aF_a^2$ is the tension of the closed domain wall. $\Gamma_s = 0.25 p\left(\Delta \theta_\pi, N_s-1\right) $ is the formation rate of seeding contours per Hubble time-space volume at e-fold $N_s$, controlled by $F_a$ during the inflationary dynamics of the ALPs, with $N_s$ being equal to 18.8 which is associated with the PTA GW signal frequency.

In numerous successful models, the inflationary energy scale $H_{\mathrm{infl}}$ typically falls within the range of approximately $10^{14}$ to $10^{16}$ GeV. The condition for wall formation necessitates $F_a>H_{\mathrm{infl}}/2\pi$, indicating a preference for a relatively low inflation scale with a fixed value of $H_{\mathrm{infl}}=10^{14}\mathrm{GeV}$. In this scenario, the only free parameter is the initial position $\theta_0$. We can identify two physically appealing and representative cases: the ``close to $\pi$'' case, where $\delta\theta_{\pi}=0.1$, and the ``far away from $\pi$'' case, where $\delta\theta_{\pi}=\pi$.

The first line in Eq.~\ref{equ:GW} corresponds to the scenario in which closed domain walls escape into baby universe, where $\kappa$ is the fraction of the radiation fluid set in motion by the sound waves. The second line pertains to the case of closed domain wall collapse into PBHs. It is important to note that there exists a critical size for the closed domain wall below which will always collapse into PBHs, resulting in a maximal PBH mass and a corresponding frequency that separates the two sources of GWs in Eq.~(\ref{equ:GW}). The critical size is determined by the moment when the energy density of the closed domain wall becomes dominant over that of the radiation fluid inside, given by the equation $\sigma R^2\simeq \rho_r R^3 = M_{\rm pl}^2 R$. Taking into account the redshift from that moment to the current time, the frequency of the turning point is given by
\begin{equation}
    \label{equ:f_peak}
    f_{\rm peak} = f \frac{a(t)}{a_0}  \simeq \sqrt{\frac{8F_a^2 m_a}{M_{\rm pl}^2t_0}}\left(\frac{\Omega_r}{\Omega_m}\right)^{1/4},
\end{equation}
where $a$ is the scale factor and $M_{\rm pl}=2.43\times10^{18}~\rm GeV$ is the reduced Planck mass. $t_0$ is the current age of the universe, and $\Omega_m$ represents the matter density of the universe. On each side of the turning frequency, the GW spectrum follows a simple power-law form, which can be compared with current observations based on PTA data.

On the other hand, the ALP closed wall network can significantly affect the primordial structure of the universe. Recently, the James Webb Space Telescope has successfully detected the galaxies at extremely high redshifts $z$ \cite{curtislake2023spectroscopic, Robertson:2022gdk}. This program employed the Near Infrared Camera to capture multi-band images ranging from $(1-5)\mathrm{\mu}\mathrm{m}$ in a designated ``blank'' field that coincided with existing Hubble Space Telescope (HST) imaging. However, the presence of six massive galaxies poses a challenge to the current theoretical frameworks of the early structure formation, as it implies a star formation efficiency, e.g., $\epsilon=0.99$ at $z=9$ and $\epsilon=0.84$ at $z=7.5$, which appears implausible in the standard $\Lambda$CDM  \cite{Boylan-Kolchin:2022kae, Lovell:2022bhx}. Nevertheless, within the regime where the star formation efficiency ranges from $0$ to $1$, the enhanced condensate may have served as the seed for primordial nonlinear structures that naturally align with the ALP closed wall.

In our proposed scenario, the formation of halos observed by JWST is governed by the ALP density contour surrounding a closed domain wall. This contour mimics the shape of the closed wall and possesses an initial energy density comparable to that of the false vacuum contained within the wall. The ALP condensate behaves as a nonrelativistic medium, leading to a decrease in energy density as the universe expands, following a proportionality to $a^{-3}$.

To derive the halo mass function for regions with enhanced ALP energy density, we utilize the mass distribution associated with the contours of the ALP closed domain wall. This distribution is determined by the number density of the seeding contours of the domain wall, which emerged during the inflationary epoch,
\begin{equation}
\frac{dn}{dM} = \frac{\Gamma_s}{2} \frac{1}{t^{3/2}} (4\pi \sigma)^{3/4} M^{-7/4},
\end{equation}
where $M$ denotes the mass of the halo. The parameter $t$ can be described as a redshift-dependent quantity using the equation $1+z = \left(1+z_{\mathrm{eq}}\right) t_{\mathrm{eq}}^{1/2} / t^{1/2}$, with $t_{\mathrm{eq}}$ representing the time of matter-radiation equality. The primary observable in our model is the Cumulative Stellar Mass Density, defined as
\begin{equation}
\rho_{\star}(M_h) = \epsilon f_b \int_{V(z_1)}^{V(z_2)} d\ln V \int_{M_h}^{\infty} \frac{dn}{d\ln M} dM,
\end{equation}
where $f_b$ is the baryon condensate enhancement factor, and the integration is performed over the desired range of redshifts from $z_1$ to $z_2$. This observable provides insights into the integrated stellar mass density within a halo of mass $M_h$ and allows us to study the relationship between the enhanced ALP energy density regions and the formation of halos and their associated stellar mass.

\section{Numerical Results and Discussions}

\begin{figure}[!htbp]
\begin{center}
\includegraphics[width=8cm]{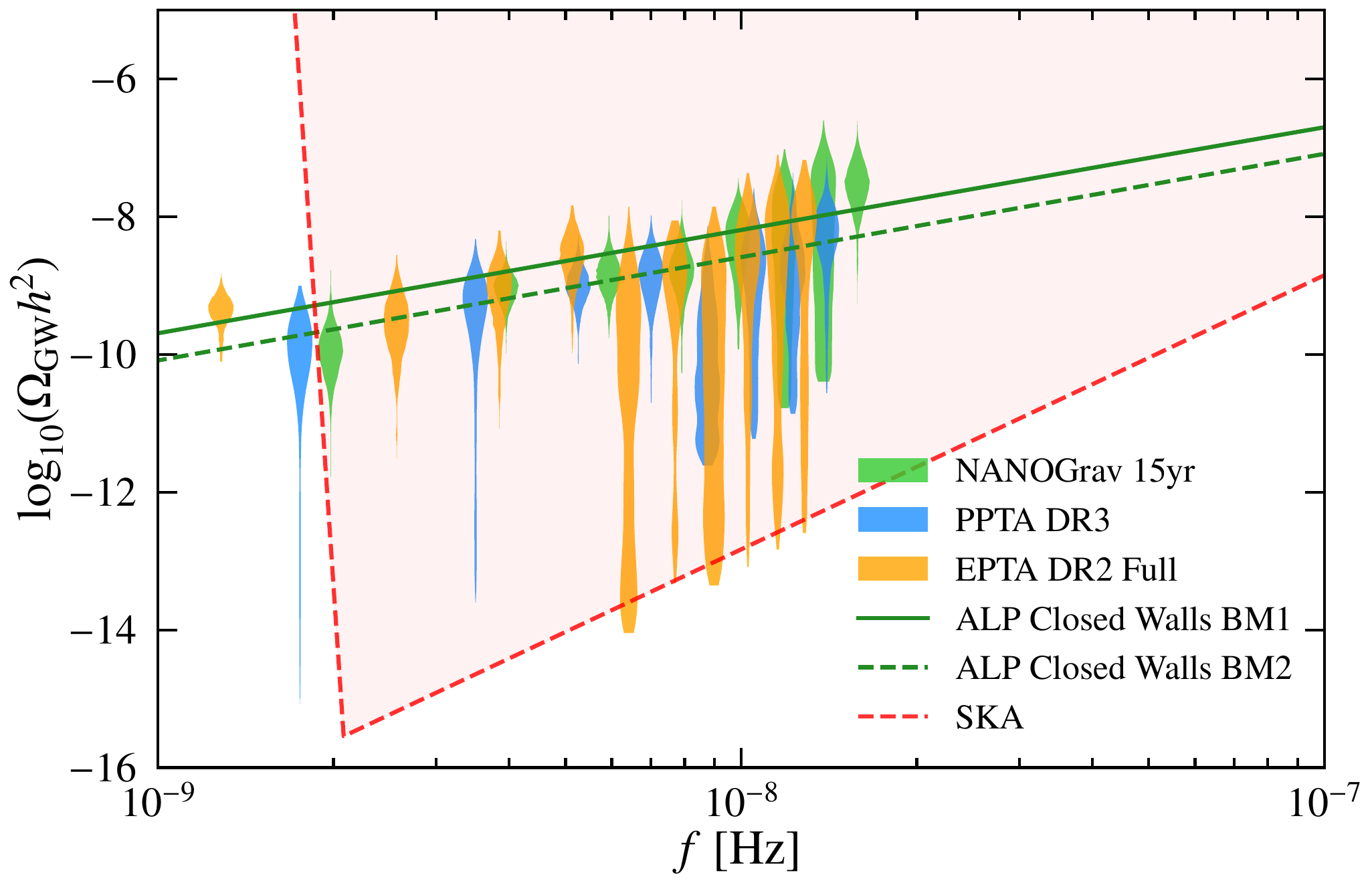}
\caption{The GW spectrum as a function of frequency. The green, blue, and orange violin plots are from the NANOGrav 15yr data~\cite{NANOGrav:2023gor}, PPTA DR3~\cite{Reardon:2023gzh} and EPTA Full DR2~\cite{Antoniadis:2023ott} respectively. The red shaded region is the projected sensitivity of the SKA~\cite{Janssen:2014dka}. The green solid and dashed lines indicate the GW spectrum from two benchmarks of the ALP closed wall network.}
\label{fig:FigOmega}
\end{center}
\end{figure}

Figure~\ref{fig:FigOmega} shows the GW power spectrum from the PTA experiments as well as the GW spectrum from the ALP closed domain wall network. The green violin plots depict the observations from 15 years NANOGrav data~\cite{NANOGrav:2023gor}, while the blue and orange violin plots show the results from PPTA third data release (DR3)~\cite{Reardon:2023gzh} and complete EPTA second data release (DR2 Full)~\cite{Antoniadis:2023ott}, respectively. Additionally, the future detection capability of SKA~\cite{Janssen:2014dka} (red) is also presented. For the GW spectrum from the ALP closed wall network, we choose two benchmarks, one for the ``far away from $\pi$'' case (BP1) and the other for the ``close to $\pi$'' case (BP2). The parameters for both benchmarks are shown in Tab.~\ref{tab:GW_BM}. The GW spectra from these two benchmarks are also shown in Fig.~\ref{fig:FigOmega}, from which, one can see that the GW spectra can be well fitted into the current PTA experiments observations. The ALP model thus offers a compelling explanation for the observed SGWB in the PTA experiments.

\begin{table}
\centering
\begin{tabular}{|l|c|c|c|c|}
\hline
 & $\delta\theta_\pi$ & $F_a$ [GeV] & $m_a$ [eV] & $\kappa$ \\
\hline
BP1 ``far away from $\pi$'' & $\pi$ & $1.0\times 10^{14}$ & \multirow{2}{*}{$10^{-2}$} & \multirow{2}{*}{$3\times10^{-5}$} \\
\cline{1-3}
BP2 ``close to $\pi$'' & $0.1$ & $3.2\times 10^{15}$ & & \\
\hline
\end{tabular}
\caption{Two benchmark points for the GW spectrum from the ALP closed wall network. One for ``far away from $\pi$'' case and one for ``close to $\pi$'' case.}
\label{tab:GW_BM}
\end{table}

In our numerical analysis, we directly fit the spectrum into the PTA observations as shown in Fig.~\ref{fig:FigOmega} by those violin plots. For NANOGrav 15yr data and PPTA DR3, we consider the first 8 bins of their results, while for EPTA Full DR2, the first 10 bins are considered. Each bin will contribute one term to the final log-likelihood:
\begin{align}
    \log\mathcal{L}_{\rm PTA} = \sum_{i=1}^8\log\mathcal{L}_i^{\rm NANOGrav} + \sum_{i=1}^{8}\log\mathcal{L}_i^{\rm PPTA} + \sum_{i=1}^{10}\log\mathcal{L}_i^{\rm EPTA}
\end{align}
where each term is calculated from the GW spectrum and the distributions from the corresponding PTA experimental results:
{
\begin{align}
    \mathcal{L}_i^{e} = \mathcal{P}_i^e(\log_{10}(\Omega_{\rm GW}(f_i^e)h^2))
\end{align}
where $\Omega_{\rm GW}(f)$ is the GW spectrum predicted from the model given in Eq.~\ref{equ:GW}. $f_i^e$ is the frequency of the i-th bin from the corresponding PTA measurements. $\mathcal{P}_i^e$ is the probability distribution in terms of $\log_{10}(\Omega_{\rm GW}h^2)$ for the i-th bin corresponding to the violin plots of each PTA measurements.
}

Assuming a generic power-law spectrum for the characteristic strain: $h_c(f)=A(f/f_{\rm yr})^{(3-\gamma)/2}$, the spectrum produced from closed walls escaping into baby universes provides $\gamma=3.5$ which dominates when the frequency is below the peak frequency in Eq.~\ref{equ:f_peak}, while those from domain wall collapse into PBH provides $\gamma=5.5$ which dominates when the frequency is above the peak frequency. However, current PTA data prefer $\gamma=3.5$ over $\gamma=5.5$. Hence, in order to simplify the analysis, it is required that the peak frequency is above the range that is covered by PTA experiments, and we only consider the contribution from the wall escaping into the baby universe $f_{\rm peak} \gtrsim f_{\rm yr}$,
where $f_{\rm yr}=1 /{\rm year}$.

\begin{figure}
    \centering
    \includegraphics[width=8cm]{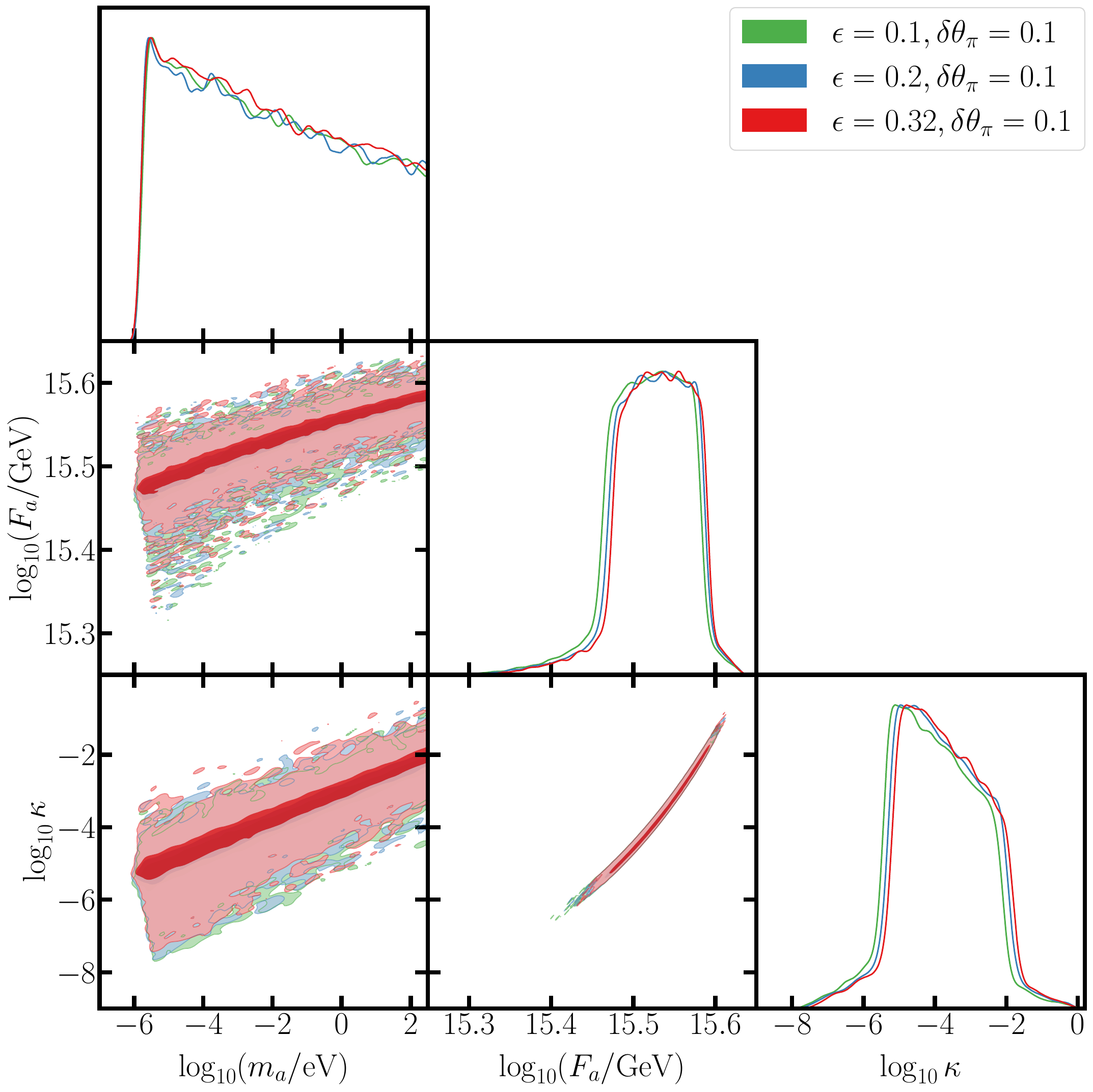}
    \includegraphics[width=8cm]{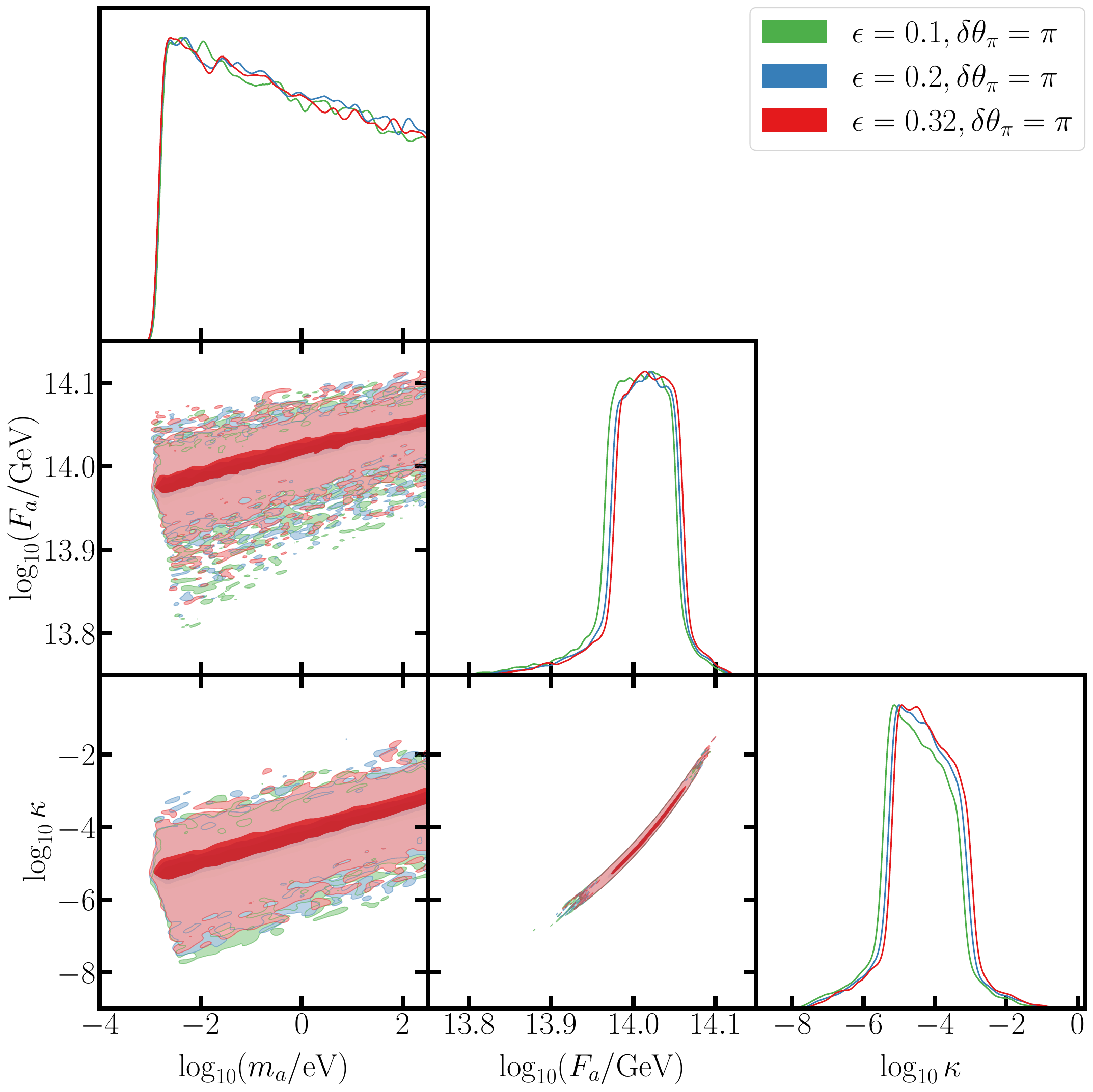}
    \caption{The 1- and 2-$\sigma$ region of the corresponding parameters under the JWST observation and PTA experiments measurements (including NANOGrav 15yr, PPTA DR3 and EPTA DR2) with three choices of $\epsilon$ for ``close to $\pi$'' ($\delta\theta_\pi=0.1$, upper panel) and ``far away from $\pi$'' ($\delta\theta_\pi=\pi$, lower panel) cases respectively. The marginalized distributions of a single variable are also shown in the diagonal figures.}
    \label{fig:JWST_PTA_MCMC}
\end{figure}

Concerning the observation from JWST, we analyze two distinct sets of data points within separate redshift bins. In the redshift range $7 < z < 8.5$, we investigate the following: $\log _{10} \rho_\star^{\text {obs }}\left(M_1\right)=5.90 \pm 0.35$ at $\log _{10}\left(M_1\right)=10.1$ and $\log _{10} \rho_{\star}^{\mathrm{obs}}\left(M_2\right)=5.70 \pm 0.65$ at $\log _{10}\left(M_2\right)=10.8$. In the redshift range $8.5 < z < 10$, we utilize the following data points: $\log _{10} \rho_{\star}^{\text {obs }}\left(M_1\right)=5.7 \pm 0.40$ at $\log _{10}\left(M_1\right)=9.7$ and $\log _{10} \rho_{\star}^{\mathrm{obs}}\left(M_2\right)=5.40 \pm 0.65$ at $\log _{10}\left(M_2\right)=10.4$. The JWST data is incorporated for each model in the chain assuming Gaussian distribution and ignoring model independent terms,
\begin{equation}
    \log\mathcal{L}_{\mathrm{JWST}}=-\frac{1}{2}\sum_{i=1}^2\left[\frac{\log _{10} \rho^{\mathrm{th}}\left(M_i\right)-\log _{10} \rho^{\mathrm{obs}}\left(M_i\right)}{\sigma_i}\right]^2
    \label{Eq:JWST}
\end{equation}
We use log-uniform distributions for the model parameters ($F_a$, $m_a$ and $\kappa$) as priors. The posteriors are obtained by implementing the total log-likelihood including the constraints from both PTA experiments and the observations from JWST in {\tt MultiNest}~\cite{Feroz:2007kg,Feroz:2008xx,Feroz:2013hea} where we choose three typical star formation efficiency $\epsilon=0.1,\,0.2,\,0.32$\footnote{Note that $\epsilon$ can also be included into the scan. However, as we will shown in the following that the dependence is weak, we hence just pick several benchmarks for $\epsilon$ for illustration.}. The results are shown in Fig.~\ref{fig:JWST_PTA_MCMC}, where the upper panel is for the ``close to $\pi$'' case while the lower panel is for the ``far away from $\pi$'' case. The darker and lighter regions (for each color) indicate the 1 and 2$\sigma$ region. The marginalized 1D distributions for each parameter are also shown in the diagonal directions. From Fig.~\ref{fig:JWST_PTA_MCMC}, we find that $F_a$ is constrained to be within a narrow region which varies according to the initial condition of the misalignment. This can be understood by examining Eq.~\ref{equ:random_walk} where large deviation from $\pi$ needs to be compensated by large fluctuations which lead to smaller $F_a$. On the other hand, a large range for $m_a$ is allowed, however, lighter ALP is preferred in the experiments. Note that the lower bound for the $m_a$ is mostly set by the requirement $f_{\rm peak} \gtrsim f_{\rm ref}$. The strong correlation between $F_a$ and $\kappa$ mainly comes from the PTA experiments. The combined results from NANOGrav, PPTA and EPTA favor only a small region for the GW spectrum amplitude. Further, either for ``close to $\pi$'' or ``far away from $\pi$'' case, the results are not sensitive to the choice of $\epsilon$, which further relaxes the current tension between the theoretical predictions and the observations for the star formation.

\begin{figure}
\centering
\includegraphics[width=8.5cm]{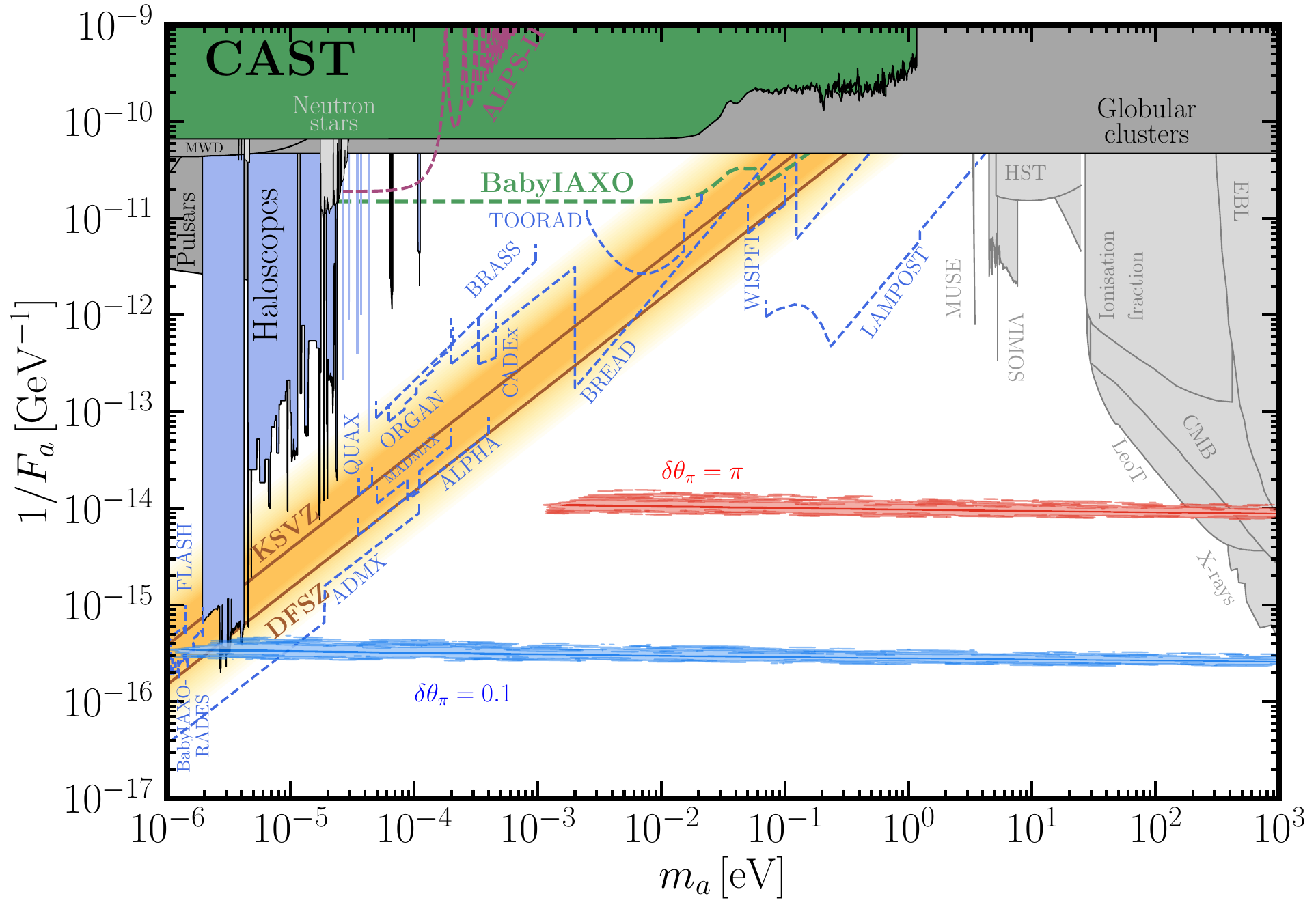}
\caption{Constraints on ALP mass and $F_a$ as well as the favored parameter region from JWST and PTA experiments (blue and red regions). The current constraints are
from CAST~\cite{CAST:2007jps,CAST:2017uph}, MUSE~\cite{Regis:2020fhw,Todarello:2023hdk}, VIMOS~\cite{Grin:2006aw}, HST~\cite{Carenza:2023qxh,Nakayama:2022jza}, Leo T gas temperature~\cite{Wadekar:2021qae}, globular clusters~\cite{Ayala:2014pea,Dolan:2022kul}, ionisation fraction, EBL, X-rays~\cite{Cadamuro:2011fd}, CMB~\cite{Capozzi:2023xie,Liu:2023nct},
as well as the haloscope experiments~\cite{DePanfilis:1987dk,Hagmann:1990tj,ADMX:2009iij,ADMX:2018gho,ADMX:2019uok,ADMX:2021nhd,HAYSTAC:2018rwy,HAYSTAC:2020kwv,HAYSTAC:2023cam,Lee:2020cfj,Jeong:2020cwz,CAPP:2020utb,Lee:2022mnc,Kim:2022hmg,Yi:2022fmn}.
In addition, the prospects from
FLASH~\cite{Alesini:2017ifp,FLASH}, BabyIAXO-RADES~\cite{Ahyoune:2023gfw},
ADMX~\cite{Stern:2016bbw}, QUAX~\cite{QUAX}, MADMAX~\cite{Beurthey:2020yuq}, ALPHA~\cite{Lawson:2019brd,Millar:2022peq}, ORGAN~\cite{McAllister:2017lkb}, CADEx~\cite{Aja:2022csb}, BRASS~\cite{BRASS},
BREAD~\cite{BREAD:2021tpx}, TOORAD~\cite{Schutte-Engel:2021bqm}, LAMPOST~\cite{Baryakhtar:2018doz}, WISPFI~\cite{Batllori:2023pap} and IAXO~\cite{IAXO:2019mpb} are also shown. The limits are drawn with the code from~\cite{AxionLimits}. {Also shown are the parameter space favored by QCD axion (yellow region).}}
\label{fig:ALP_Exps}
\end{figure}

The axion-photon coupling, conventionally represented as $1/F_a$, has undergone extensive investigation in terrestrial laboratories~\cite{DellaValle:2015xxa, OSQAR:2015qdv}, cosmological measurements~\cite{HESS:2013udx, Fermi-LAT:2016nkz}, and helioscopes~\cite{CAST:2017uph}. Figure~\ref{fig:ALP_Exps} highlights the parameter space of $m_a-F_a$. It is found that for the ``close to $\pi$'' ($\delta\theta_\pi=0.1$) case, the light ALP mass range has been touched by current haloscope experiments. While future improvement from ADMX~\cite{Stern:2016bbw} will further cover part of the parameter space. On the other hand, for the ``far away from $\pi$'' ($\delta\theta_\pi=\pi$) case, it is even below the sensitive region of future experiments. Hence, the scenario proposed in the current work opens a new direction for future ALP experiments.

\section{Conclusion}
\label{sec-3}
In this study, we have demonstrated that the detection of a stochastic gravitational wave background by Pulsar Timing Array collaborations has the potential to unveil a nontrivial cosmology involving axion-like particles. This scenario not only accounts for the observed stochastic gravitational wave background but also offers a solution to the origin of early galaxy formation as indicated by JWST observations. Through a comprehensive analysis of both JWST and NANOGrav data, we have obtained the best-fit values for the parameters associated with the axion closed wall. The predicted correlation among the stochastic gravitational wave background parameters and the distribution of enhanced regions in the axion-like particle fields, crucial for predicting early galaxy formation, offer a means to test the proposed model in the hidden sector. Additionally, the predicted mass of the axion-like particles poses a challenge for future experimental searches due to its observable effects. This discovery presents an exciting opportunity for investigating new physics, which can be thoroughly explored through a combination of cosmological, astrophysical, and experimental investigations.

\section*{Data availability}
The data generated during and/or analyzed during the current study are available from the authors on request.

\section*{Code availability}
The custom computer codes used to generate results are available from the authors on request.

\section*{Acknowledgements}
We thank A.Chaudhuri, S.G.Rubin and A.S.Sakharov for discussions. This work by L.W. and B.Z. is supported by the National Natural Science Foundation of China (NNSFC) under grants No. 12275134, No. 12147228, No. 12335005, and No. 12275232.
The research by M.K. was carried out at Southern Federal University with financial support from the Ministry of Science and Higher Education of the Russian Federation (State contract GZ0110/23-10-IF).
The work by X.L. and S.-Y.G. were supported by the NNSFC under Grant No. 12005180 and No. 12305113, by the Natural Science Foundation of Shandong Province under Grants No. ZR2020QA083, No. ZR2022QA026, and by the Project of Shandong Province Higher Educational Science and Technology Program under Grants No. 2022KJ271.

\bibliographystyle{bibsty}
\bibliography{refs}

\end{document}